\documentclass[11pt]{revtex4}
\usepackage{amssymb,epsf}
\usepackage{latexsym}
\begin{document}
\title{Extremal Myers-Perry black holes coupled to
Born-Infeld electrodynamics in five dimensions}
\author{Masoud Allahverdizadeh$^{1}$\footnote{masoud.alahverdi@gmail.com},
Jos\'e P. S. Lemos$^{1}$\footnote{joselemos@ist.utl.pt}, Ahmad
Sheykhi$^{2,3}$\footnote{asheykhi@shirazu.ac.ir}}
\address{ $^1$  Centro Multidisciplinar de Astrof\'{\i}sica -
CENTRA, Departamento de F\'{\i}sica,
Instituto Superior T\'ecnico - IST, Universidade T\'ecnica de Lisboa -
Av. Rovisco Pais 1, 1049-001 Lisboa, Portugal. \\
          $^2$  Physics Department and Biruni Observatory,
College of Sciences, Shiraz University, Shiraz 71454, Iran\\
          $^3$ Research Institute for Astronomy and Astrophysics of
Maragha (RIAAM), P.O. Box 55134-441, Maragha, Iran}

\begin{abstract}
\vspace*{1.5cm} \centerline{\bf Abstract} \vspace*{1cm}

We construct a new class of perturbative extremal charged rotating
black hole solutions in five dimensions in the Einstein--Born-Infeld
theory. We start with an extremal five-dimensional Myers-Perry black
hole seed in which the two possible angular momenta have equal
magnitude, and add Born-Infeld electrical charge keeping the
extremality condition as a constraint.  The perturbative parameter is
assumed to be the electric charge $q$ and the perturbations are
performed up to 4th order. We also study some physical properties of
these black holes. It is shown that the perturbative parameter $q$ and
the Born-Infeld parameter $\beta$ modify the values of the physical
quantities of the black holes. The solution suggests that the magnetic
moment and gyromagnetic ratio of the black hole spacetime change sign
when the Born-Infeld character of the solution starts to depart
strongly from the Maxwell limit.  We venture an interpretation for the
effect.

\end{abstract}
 \maketitle
\section{Introduction\label{Intro}}

The field of black hole solutions in higher dimensions was started in
1963 by Tangherlini \cite{Tan}. After some discussion of the old
problem of the dimensionality of space he proceed to find static
spherical symmetric solutions of the $d$ dimensional Einstein-Maxwell
equations, with $d\geq4$, generalizing thus the Schwarzschild and the
Reissner-Nordstr\"om solutions. Tangherlini also included a
cosmological term finding the corresponding de Sitter (dS) and anti-de
Sitter (AdS) $d$-dimensional black holes \cite{Tan}.

Since then, with the belief that at a more fundamental level
Einstein's theory has to be modified and extra dimensions might be
crucial in the process, static higher-dimensional black holes
solutions have been devised in gravitational theories such as
Kaluza-Klein, Gauss-Bonnet theory and its Lovelock extension,
scalar-tensor gravity, supergravity, and low-energy string theory.
Likewise, there has been attempts to modify Maxwell's electromagnetism
in order to deal in a more consistent manner with the point particle
problem and its field divergences. An important modification is given
by the Born-Infeld nonlinear electrodynamics.  Although it became less
prominent with the appearance of quantum electrodynamics and its
renormalization scheme, the Born-Infeld theory has been now occurring
repeatedly with the development of string theory, where the dynamics
of D-branes is governed by a Born-Infeld type action.  It is thus
natural to study higher dimensional black holes in these extended
gravity-electromagnetic set ups.

For the static case, several examples of higher dimensional black
holes can be given. Black holes in Lovelock gravity and Maxwell
electromagnetism were studied in \cite{banteitzadilemos}.  Black hole
solutions in Einstein-Born-Infeld gravity are less singular in
comparison with the Tangherlini-Reissner-Nordstr\"{o}m solution and
such solutions with and without a cosmological constant have been
discussed in general relativity
\cite{Garcia1,Wiltshire1,Aiello1,Tamaki1,Dey1,Gunasekaran}, and in
Gauss-Bonnet \cite{GB} and Lovelock \cite{Lov} gravities.  The
extension to the cases where the horizon has zero or negative
curvature has also been considered \cite{Cai1}. Scalar-tensor theories
of gravity coupled to Born-Infeld black hole solutions have also been
studied in \cite{yaz3}.  Attempts to find solutions in general
relativity coupled to other fields, such as dilaton, rank three
antisymmetric tensor, and Born-Infeld and others, have been performed
to construct exact solutions \cite{YI,Tam,yaz,she}.  In these theories
the dilaton field can be thought of as coming from a scalar field of
low energy limit of string theory. The appearance of the dilaton field
changes the asymptotic behavior of the solutions to be neither
asymptotically flat nor de Sitter or Anti-de Sitter.

Introducing rotation makes the search for solutions much more
difficult.  Following the example of the extension of the
Schwarzschild and Reissner-Nordstr\"om solutions to higher dimensions
in different theories, it was natural to do the same for the Kerr and
the Kerr-Newman solutions.  The generalization of the Kerr solution to
higher dimensional Einstein gravity was performed by Myers and Perry
\cite{Myer}. These Myers-Perry solutions include the non-trivial
cases of several modes of rotation due to the existence of other
rotation planes in higher dimensions.  The inclusion of a cosmological
constant on these higher dimensional solutions was performed in
\cite{many}.  Rotating black hole solutions in other gravity theories
like Gauss-Bonnet or Lovelock are not know.

Finally, the generalization of the Kerr-Newman black holes to higher
dimensional Einstein-Maxwell theory has not been found.  Likewise, the
generalization of Kerr with some other gauge field, different from
Maxwell, to higher dimensions has not had success.  In order to study
this problem one has to resort to approximations, as was done in
\cite{Aliev2}, when studying charged rotating black hole solutions in
higher dimensions with a single rotation parameter in the limit of
slow rotation (see also \cite{Aliev3,Aliev4,kunz1}).  Further study,
relying on perturbative or numerical methods, to construct those
solutions in asymptotically flat backgrounds has been performed in
\cite{Kunz2} and in asymptotically anti-de Sitter spacetimes in
\cite{Kunz3}.  Employing higher perturbation theory with the electric
charge as the perturbation parameter, solutions and properties of
charged rotating Einstein-Maxwell asymptotically flat black holes have
been constructed in five dimensions \cite{Navarro}. Focusing on
extremal black holes, this perturbative method was also applied to
obtain Einstein-Maxwell black holes with equal magnitude angular
momenta in odd dimensions \cite{Allahverdi1}.  A generalization to
include a scalar field in a Einstein-Maxwell-dilaton theory was
performed in \cite{Allahverdi2} where black holes with equal magnitude
angular momenta in general odd dimensions were obtained.

In this paper, we use this perturbative approach to find extremal
rotating Einstein-Born-Infeld black holes in a five dimensional
spacetime. Starting from the Myers-Perry black holes \cite{Myer}, we
evaluate the perturbative series up to 4th order in the Born-Infeld
electric charge parameter $q$. We determine the physical properties of
these black holes for general Born-Infeld coupling constant $\beta$.
In fact, we study the effects of the perturbative parameter $q$ and
the Born-Infeld parameter $\beta$ on the mass, angular momentum,
magnetic moment, gyromagnetic ratio, and horizon radius of these
rotating black holes. 

The outline of this paper is as follows: In section \ref{FIRST}, we
present the basic field equations of nonlinear Born-Infeld theory in
Einstein gravity and obtain a new class of perturbative charged
rotating solutions in five dimensions. In section \ref{PQ}, we
calculate the physical quantities of the solutions and discuss their
properties. In section \ref{MassForm} we study the mass formula for
these black holes.  Section \ref{SumCon} is devoted to conclusions.

\section{Metric and Gauge Potential\label{FIRST}}
We start  with the Einstein-Hilbert action coupled to the
Born-Infeld nonlinear gauge field in five dimensions
\begin{eqnarray}
S &=&\int dx^{5}\sqrt{-g}\left( \frac{R}{16 \pi G_5}\text{
}+L(F)\right),  \label{Lag}
\end{eqnarray}
where $G_5$ is five dimensional Newtonian constant, ${R}$ is the
curvature scalar and $L(F)$ is the Lagrangian of the nonlinear
Born-Infeld gauge field given by
\begin{eqnarray}
L(F) &=&4\beta^{2}\left(1-\sqrt{1+\frac{F^{\mu \nu }F_{\mu \nu
}}{2 \beta^{2}}}\right), \label{Born-Infeld}
\end{eqnarray}
where, in turn, $F_{\mu \nu }=\partial _{\mu }A_{\nu }-\partial
_{\nu }A_{\mu }$ is the electromagnetic field tensor, $A_{\mu }$
is the electromagnetic vector potential, and $\beta $ is the
Born-Infeld parameter with unit of mass. In the limit $\beta
\rightarrow \infty $, $L(F)$ reduces to the Lagrangian of the
standard Maxwell field, $L(F)={F^{\mu \nu }F_{\mu \nu }}$. The
equations of motion can be obtained by varying the action with
respect to the gravitational field $g_{\mu \nu }$ and the gauge
field $A_{\mu }$. This procedure yields the gravitational field
equations
\begin{equation}
G_{\mu \nu }=R_{\mu \nu }-\frac{1}{2}g_{\mu \nu } R=
\frac{1}{2}g_{\mu \nu } L(F)+\frac{2 F_{\mu \eta }F_{\nu }^{\text{
}\eta }}{\sqrt{1+\frac{F^{\mu \nu }F_{\mu \nu }}{2 \beta^{2}}}},
\label{FE1}
\end{equation}
and the electromagnetic equation
\begin{equation}
\partial_{\mu}{\left(\frac{\sqrt{-g}F^{\mu \nu }}
{\sqrt{1+\frac{F^{\mu \nu }F_{\mu \nu }}{2 \beta^{2}}}}
\right)}=0\,. \label{FE2}
\end{equation}
Our aim here is to find perturbative extremal charged rotating
black hole solutions of the above field equations in five
dimensions. Using coordinates $(t,r,\theta,\varphi_1,\varphi_2)$,
the $5$-dimensional Myers-Perry solution \cite{Myer} restricted to
the case where the two possible angular momenta have equal
magnitude, and following \cite{Navarro} for the parametrization of
the metric, one has
\begin{eqnarray}\label{metric1}
ds^2 &=& g_{tt} dt^2+\frac{dr^2}{W} +
r^2\left(d\theta^{2}+\sin^{2}\theta d\varphi^{2}_{1}+
\cos^{2}\theta d\varphi^{2}_{2}\right)
+N\left(\varepsilon_{1}\sin^{2}\theta d\varphi_{1}+
\varepsilon_{2}\cos^{2}\theta d\varphi_{2}\right)^{2}\nonumber \\
&&-2B\left(\varepsilon_{1}\sin^{2}\theta d\varphi_{1}+
\varepsilon_{2}\cos^{2}\theta d\varphi_{2}\right)dt,
\end{eqnarray}
where $\varepsilon_{k}$ denotes the sense of rotation in the
$k$-th orthogonal plane of rotation, such that
$\varepsilon_{k}=\pm1$, $k=1,2$. For the mentioned Myers-Perry
solution \cite{Myer} one has $g_{tt} = -1+\frac{2\hat{M}}{r^{2}}$,
$W =1-\frac{2\hat{M}}{r^{2}}+ \frac{2\hat{J}^{2}}{\hat{M}r^{4}}$,
and $N = \frac{2\hat{J}^{2}}{\hat{M}r^{2}}$, where $\hat{M}$ and
$\hat{J}$ are two constants, namely, the mass and angular momentum
parameters, respectively, related to the mass $M$ and angular
momenta $J$ of the Myers-Perry solution, through the relations
$\hat{M}=\frac{16\pi G_{5}}{3A}M$ and $\hat{J}=\frac{4\pi
G_{5}}{A}J$, where $A$ is the area of the unit $3$-sphere. An
adequate parametrization for the gauge potential is given by
\begin{eqnarray}\label{A1}
A_{\mu}dx^{\mu} &=&
a_{0}+a_{\varphi}\left(\varepsilon_{1}\sin^{2}\theta
d\varphi_{1}+\varepsilon_{2}\cos^{2}\theta d\varphi_{2}\right).
\end{eqnarray}
We further assume the metric functions $g_{tt}$, $W$, $N$, $B$ and
also the two functions $a_{0}$, $a_{\varphi}$ for the gauge field,
depend only on the radial coordinate $r$.

We consider perturbations around the Myers-Perry solution, with a
Born-Infeld 
electric charge $q$ as the perturbative parameter, so that
$q$ is much less than $\hat{M}$ or $\hat{J}^{\,2/3}$.
Taking into account
the symmetry with respect to charge reversal and the
seed solution, the metric and gauge potentials
take the form
\begin{eqnarray}\label{g_{tt}}
g_{tt} = -1+\frac{2\hat{M}}{r^{2}}+q^{2}g^{(2)}_{tt}+
q^{4}g^{(4)}_{tt}+O(q^{6}) ,
\end{eqnarray}
\begin{eqnarray}\label{W}
W =1-\frac{2\hat{M}}{r^{2}}+
\frac{2\hat{J}^{2}}{\hat{M}r^{4}}+q^{2}W^{(2)}+q^{4}W^{(4)}+O(q^{6}) ,
\end{eqnarray}
\begin{eqnarray}\label{N}
N = \frac{2\hat{J}^{2}}{\hat{M}r^{2}}+q^{2}N^{(2)}+q^{4}N^{(4)}+O(q^{6}) ,
\end{eqnarray}
\begin{eqnarray}\label{B}
B = \frac{2\hat{J}}{r^{2}}+q^{2}B^{(2)}+q^{4}B^{(4)}+O(q^{6}) ,
\end{eqnarray}
\begin{eqnarray}\label{a0}
a_{0} = q a^{(1)}_{0}+q^{3} a^{(3)}_{0}+O(q^{5}) ,
\end{eqnarray}
\begin{eqnarray}\label{aph}
a_{\varphi} = q a^{(1)}_{\varphi}+q^{3} a^{(3)}_{\varphi}+O(q^{5})\,.
\end{eqnarray}
Here $g^{(2)}_{tt}$ and
$g^{(4)}_{tt}$ are
the second and fourth order perturbative terms, respectively.
The other perturbative terms are defined similarly.

We now fix the angular momenta at any perturbative order, and
impose the extremal condition in all orders. We also assume that
the horizon is regular. With these assumptions we are able to fix
all constants of integration. To simplify the notation we
introduce a parameter $\nu$ through the equations
\begin{equation}
\hat{M}=2\nu^{2}\quad,\quad \hat{J}=2\nu^{3}\,,
\label{nu}
\end{equation}
meaning that the extremal Myers-Perry
solution holds in five dimensions. Then,
using the field equations (\ref{FE1})-(\ref{FE2}), the
perturbative solutions up to 4th order can be written as,
\begin{eqnarray}\label{gg_{tt}}
g_{tt} &=& -1+\frac{4\nu^{2}}{r^{2}}+
\frac{(r^{2}-4\nu^{2})q^{2}}{3\nu^{2}r^{4}}
+\Bigg{\{}{\frac {11}{135}}\,
{\frac {1}{{\nu}^{2}{\beta}^{2}{r}^{8}}}-{\frac {16
}{45}}\,{\frac {{\nu}^{2}}{{\beta}^{2}{r}^{12}}}+
{\frac {58}{135}}\,{
\frac {1}{{\beta}^{2}{r}^{10}}}\nonumber \\
&&-\,{\frac
{8\,{\nu}^{2}{\beta}^{2}+3}{9{\beta}^{2}{\nu}^{6}{r}^{4}}} +{\frac
{1}{540}}\,{\frac {79+240\,{\nu}^{2}{\beta}^{2}}{{\beta}^{2}{
\nu}^{4}{r}^{6}}}+{\frac {1}{720}}\,{\frac
{79+220\,{\nu}^{2}{\beta}^{
2}}{{\beta}^{2}{\nu}^{8}{r}^{2}}}\nonumber \\
&&{\frac {4 \left( {r}^{2}-2\,{\nu}^{2} \right)
^{2}}{27{\beta}^{2}{\nu}^{10}{r}^{4}}}\, \left( {\nu
}^{2}{\beta}^{2}+\frac{3}{8} \right) \ln  \left( 1-2\,{\frac
{{\nu}^{2}}{{r}^{ 2}}} \right)
 \Bigg{\}}q^{4} +O(q^{6}) ,
\end{eqnarray}
\begin{eqnarray}\label{WW}
W &=&1-\frac{4\nu^{2}}{r^{2}}+\frac{4\nu^{4}}{r^{4}}-
\frac {q^{2}(r^{2}-2\nu^{2})}{3\nu^{2}r^{4}}
\nonumber \\
&& +
\Bigg{\{}{\frac {1}{2160}}\,{\frac
{522+480\,{\nu}^{2}{\beta}^{2}}{{\beta}^{2}{
\nu}^{10}}}+{\frac {1}{2160}}\,{\frac {-2420\,{\nu}^{4}{\beta}^{2}-
2607\,{\nu}^{2}}{{\beta}^{2}{\nu}^{10}{r}^{2}}}+{\frac {1}{2160}}\,{
\frac {3840\,{\nu}^{4}+3620\,{\beta}^{2}{\nu}^{6}}{{\beta}^{2}{\nu}^{
10}{r}^{4}}}\nonumber \\
&& +{\frac {1}{2160}}\,{\frac {-1032\,{\nu}^{6}-1280\,{\nu}^{
8}{\beta}^{2}}{{\beta}^{2}{\nu}^{10}{r}^{6}}}-
\frac{1}{5}\,{\frac {1}{{r}^{8}{
\nu}^{2}{\beta}^{2}}}-{\frac {23}{45}}\,{\frac {1}{{\beta}^{2}{r}^{10}
}}+{\frac {44}{45}}\,{\frac {{\nu}^{2}}{{\beta}^{2}{r}^{12}}}-{\frac {
56}{45}}\,{\frac {{\nu}^{4}}{{\beta}^{2}{r}^{14}}}\nonumber \\
&& +\frac{\left( {r}^{2 }-2\,{\nu}^{2} \right) ^{3}}{9 {
\beta}^{2}{\nu}^{12}{r}^{4}}\,  \left( {\frac
{87}{80}}+{\nu}^{2}{\beta}^{ 2} \right) \ln  \left( 1-2\,{\frac
{{\nu}^{2}}{{r}^{2}}} \right)  \Bigg{\}}q^{4}+O(q^{6}) ,
\end{eqnarray}
\begin{eqnarray}\label{NN}
N &=& \frac{4\nu^{4}}{r^{2}}-\frac{2q^{2}(r^{2}+2\nu^{2})}{3r^{4}}+
\Bigg{\{}-{\frac {1}{360}}\,{\frac { \left( 87+80\,{\nu}^{2}{\beta}^{2}
 \right) {r}^{2}}{{\beta}^{2}{\nu}^{10}}}+{\frac {52}{135}}\,{\frac {{
\nu}^{2}}{{\beta}^{2}{r}^{10}}}+{\frac {4}{135}}\,{\frac {1}{{\beta}^{
2}{r}^{8}}}-{\frac {16}{45}}\,{\frac {{\nu}^{4}}{{\beta}^{2}{r}^{12}}}
\nonumber \\
&&+\frac{1}{45}\,
{\frac {1+20\,{\nu}^{2}{\beta}^{2}}{{r}^{6}{\nu}^{2}{\beta}^{2}
}}-{\frac {1}{90}}\,{\frac {61+40\,{\nu}^{2}{\beta}^{2}}{{\beta}^{2}{
\nu}^{4}{r}^{4}}}+{\frac {1}{360}}\,{\frac {87+80\,{\nu}^{2}{\beta}^{2
}}{{\nu}^{8}{\beta}^{2}}}+{\frac {1}{180}}\,{\frac {144+35\,{\nu}^{2}{
\beta}^{2}}{{\beta}^{2}{\nu}^{6}{r}^{2}}}\nonumber \\
&&-\,\ln  \left( 1-2\,{
\frac {{\nu}^{2}}{{r}^{2}}} \right)  \left( {\frac {87}{80}}\,{r}^{6}+
{r}^{6}{\nu}^{2}{\beta}^{2}-{\frac {57}{20}}\,{\nu}^{4}{r}^{2}+{\nu}^{
6}+\frac{8}{3}\,{\nu}^{8}{\beta}^{2} \right)
\frac{\left( {r}^{2}-2\,{\nu}^{2}
 \right)} {9{\beta}^{2}{\nu}^{12}{r}^{4}}
\Bigg{\}}q^{4}+O(q^{6}) ,
\end{eqnarray}
\begin{eqnarray}\label{BB}
B &=& \frac{4\nu^{3}}{r^{2}}-
\frac{4\nu}{3r^{4}}q^{2}+
\Bigg{\{}{\frac {1}{1440}}\,
{\frac {480\,{\nu}^{2}{\beta}^{2}+864}{{\beta}^{2}{
\nu}^{7}{r}^{2}}}+{\frac {2}{27}}\,{\frac {1}{{\beta}^{2}\nu\,{r}^{8}}
}+{\frac {52}{135}}\,{\frac {\nu}{{\beta}^{2}{r}^{10}}}-{\frac {1}{60}
}\,{\frac {29+40\,{\nu}^{2}{\beta}^{2}}{{\beta}^{2}{\nu}^{5}{r}^{4}}}
\nonumber \\
&&+
{\frac {1}{90}}\,{\frac {9+40\,{\nu}^{2}{\beta}^{2}}{{\beta}^{2}{\nu}^
{3}{r}^{6}}}-{\frac {16}{45}}\,{\frac {{\nu}^{3}}{{\beta}^{2}{r}^{12}}
}-\frac{1}{45}\,{\frac {9+5\,{\nu}^{2}{\beta}^{2}}{{\beta}^{2}{\nu}^{9}}}+
 ( -{\frac {1}{270}}\,{\frac {27+15\,{\nu}^{2}{\beta}^{2}}{{\beta
}^{2}{\nu}^{11}}}\nonumber \\
&&+{\frac {1}{270}}\,{\frac {54\,{\nu}^{2}+30\,{\nu}^{4
}{\beta}^{2}}{{\beta}^{2}{\nu}^{11}{r}^{2}}}-{\frac {1}{270}}\,{\frac
{80\,{\beta}^{2}{\nu}^{6}+30\,{\nu}^{4}}{{\beta}^{2}{\nu}^{11}{r}^{4}}
} )  \left( {r}^{2}-2\,{\nu}^{2} \right) \ln  \left( 1-2\,{
\frac {{\nu}^{2}}{{r}^{2}}} \right)\Bigg{\}}q^{4}+O(q^{6}) ,
\end{eqnarray}
\begin{eqnarray}\label{a00}
a_{0} &=& \frac{q}{r^{2}}+\Bigg{\{}{\frac {1}{180}}\,
{\frac {40\,{\nu}^{2}{\beta}^{2}+15}{{\beta}^{2}{\nu
}^{6}{r}^{2}}}-{\frac {1}{180}}\,{\frac {15\,{\nu}^{2}+40\,{\nu}^{4}{
\beta}^{2}}{{\beta}^{2}{\nu}^{6}{r}^{4}}}-{\frac {11}{180}}\,{\frac {1
}{{r}^{6}{\nu}^{2}{\beta}^{2}}}-{\frac {26}{45}}\,{\frac {1}{{\beta}^{
2}{r}^{8}}}+\frac{2}{3}\,{\frac {{\nu}^{2}}{{\beta}^{2}{r}^{10}}}\nonumber \\
&&+\frac{1}{9}\,
 \left( {\nu}^{2}{\beta}^{2}+\frac{3}{8} \right)
\left( {r}^{2}-2\,{\nu}^{2}
 \right) \ln  \left( 1-2\,{\frac {{\nu}^{2}}{{r}^{2}}} \right) {\beta}
^{-2}{\nu}^{-8}{r}^{-2}
\Bigg{\}}q^{3}+O(q^{5})  ,
\end{eqnarray}
\begin{eqnarray}\label{aphiphi}
a_{\varphi}&=&-\frac{\nu q}{r^{2}}+\Bigg{\{}-{\frac {1}{720}}\,
{\frac {80\,{\nu}^{2}{\beta}^{2}+30}{{\nu}^{7}{
\beta}^{2}}}-{\frac {1}{720}}\,{\frac {40\,{\nu}^{4}{\beta}^{2}+27\,{
\nu}^{2}}{{\nu}^{7}{\beta}^{2}{r}^{2}}}-{\frac {1}{720}}\,{\frac {-160
\,{\beta}^{2}{\nu}^{6}-88\,{\nu}^{4}}{{\nu}^{7}{\beta}^{2}{r}^{4}}}+{
\frac {7}{60}}\,{\frac {1}{\nu\,{\beta}^{2}{r}^{6}}}\nonumber \\
&&+{\frac {26}{45}}
\,{\frac {\nu}{{\beta}^{2}{r}^{8}}}-\frac{2}{3}\,
{\frac {{\nu}^{3}}{{\beta}^{2
}{r}^{10}}}-{\frac {1}{144}}\, \left( 3+8\,{\nu}^{2}{\beta}^{2}
 \right)  \left( \frac{r^{2}}{\beta^{2}\nu^{9}}-
\frac{4}{r^{2}\beta^{2}\nu^{5}} \right) \ln  \left( 1-2\,{
\frac {{\nu}^{2}}{{r}^{2}}} \right) \Bigg{\}}q^{3}+O(q^{5}).
\end{eqnarray}
It is seen that there are the usual $1/r^n$ polynomial expressions
as well as terms involving logarithmic functions. It is also worth
mentioning that the Born-Infeld parameter $\beta$ appears in terms
which are of 3rd and 4th order in the electric charge parameter
$q$. One may note that in Maxwell's limit, $\beta \longrightarrow
\infty $, these perturbative solutions reduce to the five
dimensional perturbative charged rotating black holes in
Einstein-Maxwell theory presented in \cite{Navarro}. A consistent
check of these solutions can be provided by Smarr's formula.
\section{Physical Quantities\label{PQ}}
The mass $M$, the angular momenta $J$, the electric charge $Q$, and the
magnetic moment $\mu_{\rm mag}$ can be read off the asymptotic
behavior of the metric and the gauge potential \cite{Kunz2}.
The asymptotic forms are,
\begin{eqnarray}\label{quantities}
g_{tt}=-1+\frac{\tilde{M}}{r^{2}}+...,
\quad B=\frac{2\tilde{J}}{r^{2}}+...,
 \quad  a_{0}=\frac{\tilde{Q}}{r^{2}}+...,  \quad
a_{\varphi}=\frac{\tilde{\mu}_{\rm mag}}{r^{2}}+...,
\end{eqnarray}
where $\tilde{M}$, $\tilde{J}$, $\tilde{Q}$, and $\tilde{\mu}_{\rm
mag}$ are the mass, angular momentum, electric charge, and
magnetic moment parameters, respectively, and we have defined
$\tilde{Q}\equiv q$ for notational consistency. These parameters
are related to the real mass $M$, angular momentum $J$, electric
charge $Q$, and magnetic moment ${\mu}_{\rm mag}$, through the
relations,
\begin{eqnarray}\label{quantities1}
\tilde{M}&=&\frac{16\pi G_{5}}{3A}M, \ \  \quad
\tilde{J}=\frac{4\pi G_{5}}{A}J,\nonumber \\ \quad
\tilde{Q}&=&\frac{4\pi G_{5}}{2A}Q, \quad \tilde{\mu}_{\rm
mag}=\frac{4\pi G_{5}}{2A}\mu_{\rm mag},
\end{eqnarray}
Note that when the perturbative parameter $q$ is equal to zero,
the tilde quantities of Eq.~(\ref{quantities}) reduce to the hat
quantities of Eq.~(\ref{nu}), i.e., $\tilde{M}=\hat{M}$,
$\tilde{J}=\hat{J}$, $\tilde{Q}=\hat{Q}=0$, and $\tilde{\mu}_{\rm
mag}=\hat{\mu}_{\rm mag}=0$.

Now, comparing the expansions in Eqs.~(\ref{quantities}) and
Eq.~(\ref{quantities1}) with the asymptotic behavior of the
solutions given in Eqs.~(\ref{gg_{tt}})-(\ref{aphiphi}), we obtain
\begin{eqnarray}\label{mass}
M =\frac{3\pi\nu^{2}}{2}+\frac{\pi q^{2}}{8\nu^{2}}+
\frac{\pi q^{4}
\left(20\nu^{2}\beta^{2}-3\right)}{5760\beta^{2}\nu^{8}}+O(q^{6}) ,
\end{eqnarray}
\begin{eqnarray}\label{ang-mom}
J =\pi \nu^{3} ,
\end{eqnarray}
\begin{eqnarray}\label{charge}
Q =\pi q ,
\end{eqnarray}
\begin{eqnarray}\label{mag-mom}
\mu_{\rm mag} = \pi \nu q-\frac{\pi
q^{3}\left(40\nu^{2}\beta^{2}+3\right)}{720\nu^{5}\beta^{2}}+O(q^{5}),
\end{eqnarray}
The gyromagnetic ratio $g$ is then given by
\begin{eqnarray}\label{g}
g = 2\, \frac{\mu_{\rm mag}/Q}{J/M}=\frac{2M\mu_{\rm mag}}{QJ}
=3+\frac{q^{2}
\left(20\nu^{2}\beta^{2}-3\right)}{240\nu^{6}\beta^{2}}-\frac{q^{4}
\left(10\nu^{2}\beta^{2}+3\right)}{1440\nu^{10}\beta^{2}}+O(q^{6})\,.
\end{eqnarray}
The horizon radius $r_H$ is given by
\begin{eqnarray}\label{rh}
r_H=\sqrt {2}\nu+{\frac {{q}^{2}\sqrt {2}}{24{\nu}^{3}}}+{\frac {11}{
1152}}\,{\frac {{q}^{4}\sqrt {2}}{{\nu}^{7}}}
+O(q^{6})\,.
\end{eqnarray}
All these quantities are worth commenting.

The mass $M$ of the black holes as a function of the Born-Infeld
parameter $\beta$ has an interesting behavior, as shown in Fig.~1,
see also Eq.~(\ref{mass}).  The mass $M$ increases with increasing
$\beta$ and as $\beta\rightarrow\infty$, i.e., in the Maxwell
limit, the mass takes the value $M
=\frac{3\pi\nu^{2}}{2}+\frac{\pi q^{2}}{8\nu^{2}}+\frac{\pi
q^{4}}{288\nu^{6}}+O(q^{6})$, exactly the result obtained for the
five-dimensional perturbative Einstein-Maxwell black hole
\cite{Navarro}.  For very small $\beta$ the mass turns negative.
We do not attach any significance to this result since the value
of $\beta$ for which the mass is zero uses a perturbative $q^2$
term which is much larger than than the zeroth order term in the
expressions for the magnetic moment and gyromagnetic ratio.  In
fact, the values of $\beta$ for which the results make thorough
sense are values of $\beta$ larger than the ones which yield the
zeros of the magnetic moment and gyromagnetic ratio.

The angular momenta $J$ in Eq.~(\ref{ang-mom}) and the charge $q$
in Eq.~(\ref{charge}) are fixed and do not depend on $\beta$,
following our approach. The magnetic dipole moment $\mu_{\rm mag}$
given in Eq.~(\ref{mag-mom}) appears due to the rotation of the
electrically charged black hole. The first order term
$\pi\,q\,\nu\ =Q\,\nu$ is equivalent to the magnetic moment of a
point particle rotating around an axis, with charge $q_{\rm
particle}=3\pi Q$, and with the same angular momentum $j_{\rm
particle}$ and mass $m_{\rm particle}$ of the black hole, i.e.,
$j_{\rm particle}=J$ and $m_{\rm particle}=M$, since for such a
system $\mu_{\rm mag}= \frac12\,q_{\rm particle} \frac{j_{\rm
particle}}{m_{\rm particle}}= Q\,\nu$. The higher order terms
presumably come from the spacetime curvature.  From Fig.~2 we find
out that the magnetic dipole moment $\mu_{\rm mag}$ increases with
increasing $\beta$.  In the limit $\beta \rightarrow \infty$, the
mass, the magnetic moment $\mu_{\rm mag} = \pi \nu q-\frac{\pi
q^{3}}{18\nu^{3}}+O(q^{5})$ which is exactly the result obtained
for the five-dimensional perturbative Einstein-Maxwell black hole
\cite{Navarro}. For small $\beta$ the magnetic moment is negative.
We analyze this result below, when we comment on the gyromagnetic
ratio.

\begin{figure}[tbp]
\epsfxsize=6cm \centerline{\epsffile{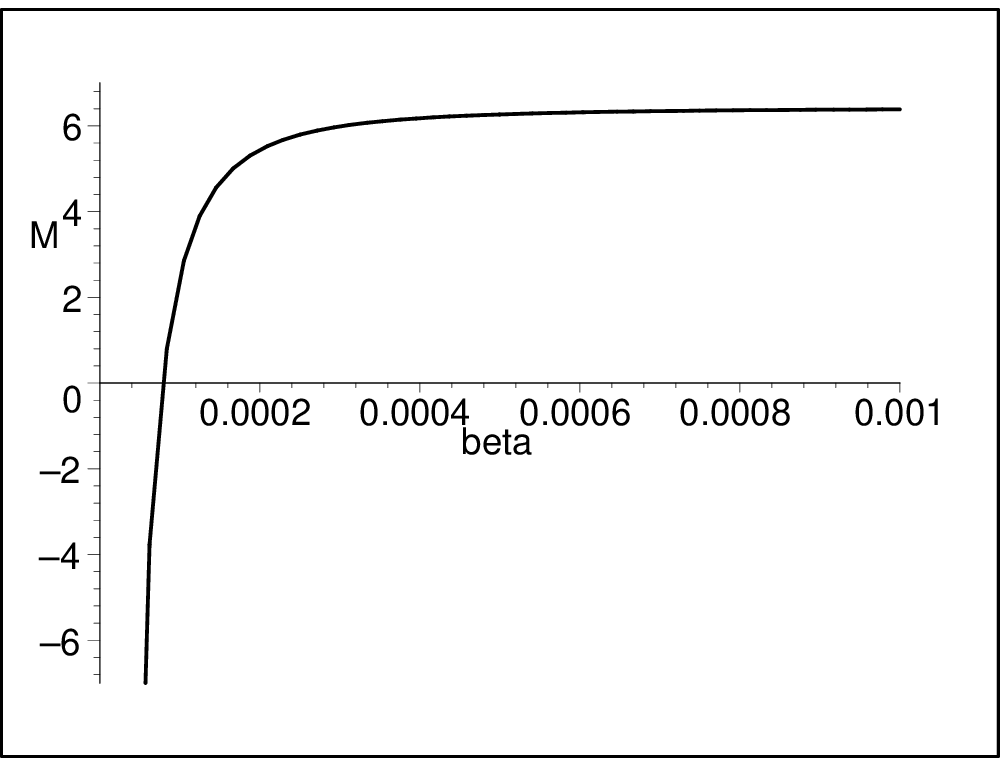}} 
\vskip -0.5cm
\caption{The black hole mass
$M$ versus the Born-Infeld parameter $\beta$ for $\nu=1.16$ and
$q=0.09$.} \label{fig1}
\vskip 1.1cm
\epsfxsize=6cm \centerline{\epsffile{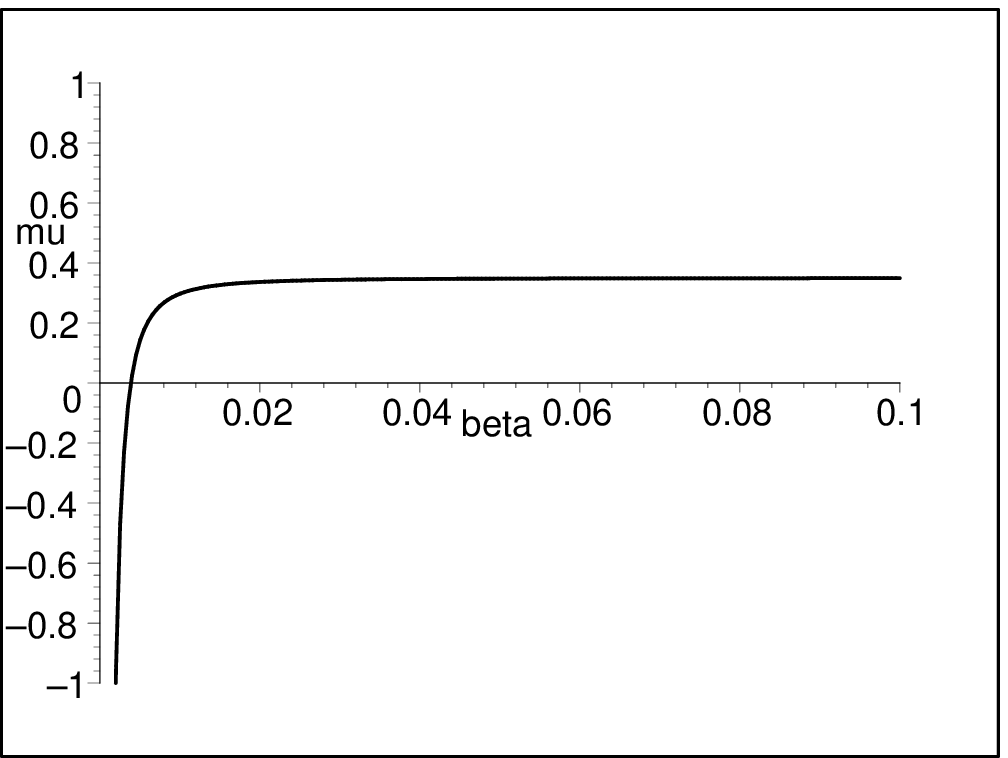}}  
\vskip -0.5cm
\caption{The black hole
magnetic moment $\mu_{\rm mag}$ versus the Born-Infeld parameter
$\beta$ for $\nu=1.16$ and $q=0.09$.} \label{fig2}
\vskip 1.1cm
\epsfxsize=6cm \centerline{\epsffile{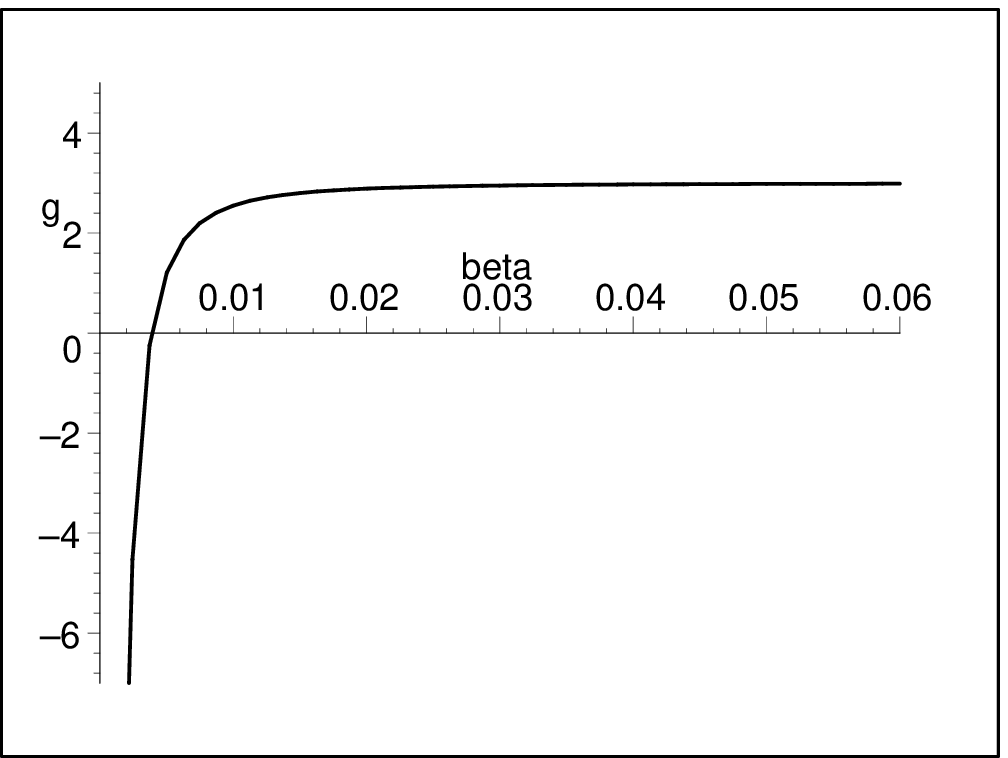}}  
\vskip -0.5cm
\caption{The black hole
gyromagnetic ratio $g$ versus the Born-Infeld parameter $\beta$ for
$\nu=1.16$ and $q=0.09$.} \label{fig3}
\end{figure}

The dimensional gyromagnetic ratio for a given system defined in
Eq.~(\ref{g}) is twice the ratio of the magnetic moment divided by
the charge to the angular momentum divided by the mass.  It has
the value $1$ for a classical body with uniform mass and uniform
charge distribution rotating about an axis of symmetry.  For an
electron it has the value of $2.00$ plus small quantum
corrections, and for the proton $5.59$ and neutron has the value
$5.59$ and $-3.826$, respectively.  From Eq.~(\ref{g}) we see that
the perturbative parameter $q$ and the parameters $\beta$ and
$\nu$ modify the gyromagnetic ratio of asymptotically flat
five-dimensional charged rotating black holes as compared to the
uncharged extremal Myers-Perry black holes. We want to study in
more detail this modification of the gyromagnetic ratio when one
varies the Born-Infeld parameter $\beta$, see Fig.~3. From the
figure we find that the gyromagnetic ratio $g$ increases with
increasing $\beta$, and in the limit $\beta \rightarrow \infty$,
the gyromagnetic ratio reduces to $g
=3+\frac{q^{2}}{12\nu^{4}}-\frac{q^{4}}{144\nu^{8}}+O(q^{6})$
which is exactly the result obtained for the five-dimensional
perturbative Einstein-Maxwell black hole \cite{Navarro}.  Now,
from Fig.~3 one finds that for some low value of $\beta$ the
gyromagnetic ratio is zero, and then turns negative. This change
of sign comes from a perturbative $q^2$ term, and thus the result
might not hold for the full exact solution. On the other hand the
result is sufficiently intriguing that deserves some attention.
One can speculate that it is at least qualitatively correct, and
perhaps expected. Let us see why.  It is known that the
Born-Infeld theory is different from Maxwell's theory when the
electromagnetic fields are very strong. The Born-Infeld theory
gives a finite total energy $E$ for the field around a point
particle with charge $q_{\rm particle}$, indeed $E\simeq
\sqrt{q_{\rm particle}^3\,\beta}$.  It also gives an effective
radius $r_0$ for the charge distribution, $r_0=\sqrt{\frac{q_{\rm
particle}}{\beta}}$. Curiously, the reversal of the gyromagnetic
ratio $g$ in Fig.~3 (concomitant to the reversal of the magnetic
dipole moment $\mu_{\rm mag}$ in Fig.~2) happens when
$r_0=\sqrt{\frac{Q}{\beta}}$ is of the order or larger than the
horizon radius $r_H$.  Indeed, for the values of $\nu$ and $q$
used in the figures, one finds $r_0=8.79$ and $r_H=1.65$.  This
reversal could then be interpreted as follows.  For large $\beta$
much of the electrical charge is distributed in a point-like
manner, as in Maxwell's theory. For small $\beta$ the charge
distribution is extended, and for sufficiently small $\beta$ it
even extends outside the horizon.  It is known that an object with
magnetic moment ${\vec\mu}_{\rm mag}$ placed on a magnetic field
$\vec{B}$ suffers a torque given by ${\vec\mu}_{\rm mag}\times
\vec{B}$. A black hole with magnetic moment is also subjected to
this kind of torque.  So, for large $\beta$, i.e., the point like
case, when a magnetic field is applied to the black hole spacetime
the resultant torque on the black hole tends to rotate it in the
expected sense, and thus the magnetic dipole moment $\mu_{\rm
mag}$ and $g$ are positive, On the other hand, when $\beta$ is
small, i.e., the charge distribution outside the black hole case,
it is the region external to the black hole horizon that is
effectively charged and it is this very region that upon
application of a magnetic field tends to rotate in the expected
sense. So here, the black hole rotates in the opposite sense,
giving a negative magnetic dipole moment $\mu_{\rm mag}$ and thus
a negative gyromagnetic ratio $g$.

The horizon radius $r_H$ in Eq.~(\ref{rh}) has the unexpected
feature that it does not depend on $\beta$ at least up to 4th
order in the charge. Up to this order $r_H$ is equal to the
Einstein-Maxwell case \cite{Navarro}.

\section{The mass formula\label{MassForm}}
Define $\xi$ as the timelike Killing vector and $\eta_{k}$,
$k=1,2$, as the two azimuthal Killing vectors. The two equal
horizon constant angular velocities $\Omega$ can then be defined
by imposing that the Killing vector field
\begin{eqnarray}\label{chi}
\chi = \xi+\Omega\sum^{2}_{k=1}\epsilon_{k}\eta_{k},
\end{eqnarray}
is null on the horizon and orthogonal to it as well.
This yields,
\begin{eqnarray}\label{Omega}
\Omega = \frac{1}{2\nu}-\frac{q^2}{24\nu^{5}}-
\frac{q^{4}
\left(5\nu^{2}\beta^{2}-1\right)}{1440\beta^{2}\nu^{11}}+O(q^{6})\,.
\end{eqnarray}
The 3-area of the horizon $A_{H}$ and the electrostatic potential
at the horizon $\Phi_{H}$ are given by
\begin{eqnarray}\label{AH}
A_{H} =8\pi^{2}\nu^{3}+O(q^{6})\, .
\end{eqnarray}
\begin{eqnarray}\label{PhiH}
\Phi_{H} =\frac{q}{4\nu^{2}}+\frac{q^{3}
\left(20\beta^{2}\nu^{2}\right)}{1440\beta^{2}\nu^{8}}+O(q^{5})\,.
\end{eqnarray}
The surface gravity $\kappa$ is  defined by $
\kappa^{2}=-\frac{1}{2}(\nabla_{\mu}
\chi_{\nu})(\nabla^{\mu}\chi^{\nu}) $. Taking into account the
conserved quantities obtained in the last section,  one can check
that these quantities satisfy the Smarr mass formula up to 4th
order \cite{Gunasekaran}. Indeed, in general the formula is
\begin{eqnarray}
 M =
\frac{3\kappa_{sg} A_{H}}{16\pi G_{D}}+3\Omega
J+\Phi_{H}Q-\frac{\beta }{2}\frac{\partial M}{\partial \beta}.
\end{eqnarray}
For an extremal solution with $\kappa=0$, the Smarr mass formula
reduces to
\begin{eqnarray}\label{smarr}
M = 3\Omega
J+\Phi_{H}Q-\frac{\beta }{2}\frac{\partial M}{\partial \beta}\,.
\end{eqnarray}
Taking into account that the mass $M$ of the black holes is given
by  Eq.~(\ref{mass}), one can determine the last term in
(\ref{smarr}), and find
\begin{eqnarray}\label{smarr2}
M = 3\Omega
J+\Phi_{H}Q+\Phi_{\beta}Q^4,
\end{eqnarray}
where
\begin{eqnarray}\label{Phibeta}
\Phi_{\beta}=-\frac{\beta }{2Q^4} \frac{\partial M}{\partial
\beta}=-{\frac {1}{1920}}\, {\frac {1}{{\pi
}^{3}{\nu}^{8}{\beta}^{2}}}\,.
\end{eqnarray}
\section{Conclusions\label{SumCon}}
In conclusion, we have presented a new class of perturbative
charged rotating black hole solutions in five dimensions in the
presence of a nonlinear Born-Infeld gauge field. This class of
solutions is restricted to the extremal black holes with equal
angular momenta. At infinity, the metric is asymptotically locally
flat. Our strategy for obtaining these solutions was through a
perturbative method up to the $4$th order for the perturbative
parameter $q$. We have started from rotating Myers-Perry black
hole solutions \cite{Myer} in five dimensions, and then studied
the effects of adding a charge parameter to the solutions. We have
calculated the conserved quantities of the solutions such as mass,
angular momentum, electric charge, magnetic moment, gyromagnetic
ratio, and horizon radius.  We found that the Born-Infeld
parameter $\beta$ modifies the values of all the physical
quantities, except the horizon radius, relative to the
corresponding Einstein-Maxwell five dimensional rotating
solutions. For large $\beta$ the solutions reduce to the
perturbative rotating Einstein-Maxwell solutions \cite{Navarro},
as we expected. We also speculated on what might happen for these
solutions in the strong electromagnetic regime, i.e., when $\beta$
is small. The generalization of the present work to all higher
dimensional is quite an interesting subject which will be
addressed elsewhere.
\acknowledgments{The support of the Funda\c{c}\~{a}o para a
Ci\^{e}ncia e a Tecnologia of Portugal Project
PTDC/FIS/098962/2008 and PEst-OE/FIS/UI0099/2011 is gratefully
acknowledged.  M. Allaverdizadeh is supported by a FCT grant.  The
work of A. Sheykhi has been supported financially by Research
Institute for Astronomy and Astrophysics of Maragha (RIAAM), Iran.

\end{document}